\documentclass[aps,prl,twocolumn,floatfix,showpacs]{revtex4}
\usepackage[dvips]{graphicx}
\usepackage{amssymb,amsfonts,amsmath,longtable}

\begin{document}

\title{Cages and anomalous diffusion in vibrated dense granular media}

\author{Camille Scalliet}
\affiliation{Universit\'e de Lyon, Ecole Normale Sup\'erieure de Lyon, Laboratoire de physique, 46 All\'ee d'Italie, 69364 Lyon, Cedex 07, France}
\affiliation{Dipartimento di Fisica, Universit\`a ``Sapienza'', Piazzale Aldo Moro 5, 00185 Rome, Italy}

\author{Andrea Gnoli}
\affiliation{Istituto dei Sistemi Complessi - Consiglio Nazionale delle Ricerche, Rome, Italy}
\affiliation{Dipartimento di Fisica, Universit\`a ``Sapienza'', Piazzale Aldo Moro 5, 00185 Rome, Italy}

\author{Andrea Puglisi}
\affiliation{Istituto dei Sistemi Complessi - Consiglio Nazionale delle Ricerche, Rome, Italy}
\affiliation{Dipartimento di Fisica, Universit\`a ``Sapienza'', Piazzale Aldo Moro 5, 00185 Rome, Italy}

\author{Angelo Vulpiani}
\affiliation{Dipartimento di Fisica, Universit\`a ``Sapienza'', Piazzale Aldo Moro 5, 00185 Rome, Italy}
\affiliation{Istituto dei Sistemi Complessi - Consiglio Nazionale delle Ricerche, Rome, Italy}

\begin{abstract}
A vertically shaken granular medium hosts a blade rotating around a
fixed vertical axis, which acts as a mesorheological probe. At 
high densities, independently from the shaking intensity, the blade's
dynamics show strong caging effects, marked by transient
sub-diffusion and a maximum in the velocity power density spectrum
(vpds), at a resonant frequency $\sim 10$ Hz. Interpreting the data through
a diffusing harmonic cage model allows us to retrieve the elastic
constant of the granular medium and its collective diffusion
coefficient. For high frequencies $f$, a tail $\sim 1/f$  in the vpds
reveals non-trivial correlations in the intra-cage micro-dynamics. At
very long times (larger than $10$ s), a super-diffusive behavior emerges, ballistic in the
most extreme cases. Consistently, the distribution of slow velocity
inversion times $\tau$ displays a power-law decay, likely due to
persistent collective fluctuations of the host medium.
\end{abstract}


%45.70.-n	Granular systems
%47.57.Gc	Granular flow
%05.40.-a	Fluctuation phenomena, random processes, noise, and Brownian motion

\maketitle

%%%%%%%%%%%%%%%%%%%%%%%%%%%%%%%%%%%%%%%%%%%%%%%%%%%%%%%%%%%%%
%%%%%%%%%%%%%%%%%%%%%%%%%%%%%%%%%%%%%%%%%%%%%%%%%%%%%%%%%%%%%
%%%%%%%%%%%%%%%%%%%%%%%%%%%%%%%%%%%%%%%%%%%%%%%%%%%%%%%%%%%%%

A comprehensive theory of dense granular media is still
lacking~\cite{andreotti}. Unperturbed granular systems may support
external forces without flowing, behaving like a solid: in this
configuration, a granular packing may show an elastic response to
small stresses. Under gentle tapping, the granular medium undergoes
very slow rearrangements, which resemble the sluggish response of
molecular glasses~\cite{bagnold,hecke2,makse1,makse2}. The jamming transition,
observed when reducing the vibration intensity, or increasing the
density, is commonly compared to the glass transition in undercooled
liquids~\cite{hecke}.  When the energy of the external vibration is
increased or the density is decreased, the granular medium enters a
liquid-like phase~\cite{danna} which has not received as much amount
of attention as the glass/solid phases or the much more dilute gas
phase~\cite{poeschel}. Nevertheless, learning from molecular fluids,
the dynamics of the liquid phase is rich of information: in
particular, it gives important hints about the many timescales
developing when the glass transition is approached from
above~\cite{dyre}. Recent theoretical and experimental insights in the
liquid phase have highlighted the differences between molecular
glass-formers and driven granular media, in particular the presence of
superdiffusive behavior~\cite{barrat,cipelletti,bouchaud,behringer}.

As rheometers are usually conceived to apply the excitation at the
boundaries, standard rheology in granular liquids is typically limited
by problems of slip and shear localization. Other techniques exist to
probe the bulk response properties of granular media, e.g. magnetic
resonance imaging, X-ray tomography and high-speed particle
tracking~\cite{mueth}, confocal microscopy~\cite{brujic} or
multispeckle dynamic light
scattering~\cite{cipelletti}. Microrheology~\cite{squires} is also
used, e.g. to probe transient subdiffusive behavior and cage
effects~\cite{marty}. However, it is not evident the ability of small
intruders to probe the collective behavior at large spatial scales in
the host granular medium.

\begin{figure}[htbp]
\includegraphics[width=6.5cm,clip=true]{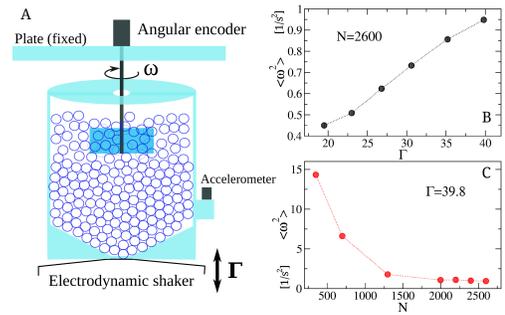}
\caption{{A: Schematic of the experimental setup. B and C: mean squared angular
velocity of the blade in the various experiments.} \label{fig:setup}} 
\end{figure}

We propose a technique which could be named passive {\em
  mesorheology}, see Fig.~\ref{fig:setup}A, inspired from previous
works~\cite{danna,danna2,noiplos,naert}. The granular medium made of
spheres of diameter $d=4$ mm is placed in a cylindrical container of
volume $\sim 7300$ times that of a sphere. The container is vertically
shaken with a signal whose spectrum is approximately flat in a range
$[f_{min},f_{max}]$ with $f_{min}=200$ Hz and $f_{max}=400$ Hz.  A
blade, our probe with cross section $\sim 8d \times 4d$, is suspended
into the granular medium and rotates around a vertical axis. Its
angular velocity $\omega(t)$ and the traveled angle of rotation
$\theta(t)=\int_0^t \omega(t')dt'$ are measured with a time-resolution
of $2$ kHz. The blade, interacting with the spheres, performs a motion
qualitatively similar to an angular Brownian motion. Two families of
experiments have been performed: a) a series at high density
($N=2600$), varying the shaking intensity $\Gamma=\ddot{z}_{max}/g
\in[19.5,39.8]$, and b) a series at high shaking intensity
($\Gamma=39.8$), varying $N \in [300,2600]$ and the packing fraction $\phi$ as indicated in the Figures. Fig.~\ref{fig:setup}B and C
report the values of the mean squared angular velocity $\langle
\omega^2 \rangle$ of the blade in the different experiments.  Details
on dimensions of the setup and shaken parameters are reported in
Supplemental Material~\cite{sm}.

%%%%%%%%%%%%%%%%%%%%%%%%%%%%%%%%%%%%%%%%%%%%%%%%%%%%%%%%%%%%%
%%%%%%%%%%%%%%%%%%%%%%%%%%%%%%%%%%%%%%%%%%%%%%%%%%%%%%%%%%%%%
%%%%%%%%%%%%%%%%%%%%%%%%%%%%%%%%%%%%%%%%%%%%%%%%%%%%%%%%%%%%%

\begin{figure}
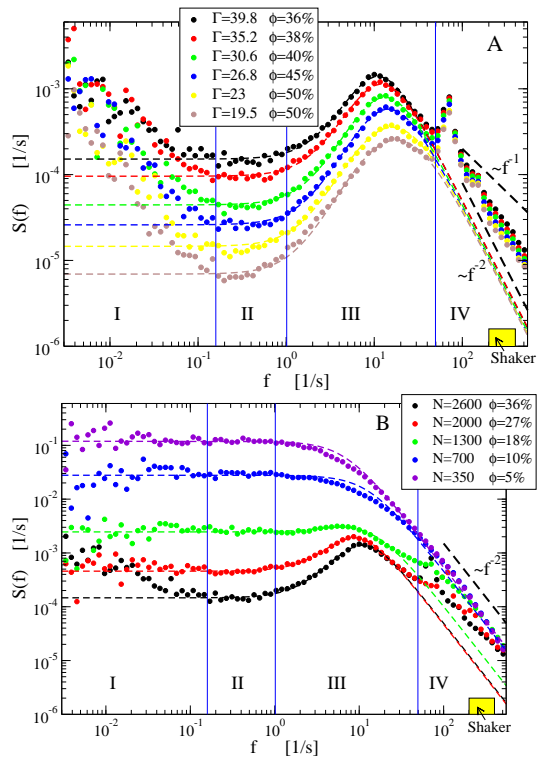

\includegraphics[width=7cm,clip=true]{spettri_varioA}
\includegraphics[width=7cm,clip=true]{spettri_varioN}
\caption{
Power density spectra of the blade's angular velocity for the two
series. A: the number of beads is fixed ($N = 2600$) and the shaking
intensity varies. B: the shaking intensity
is fixed to $\Gamma=39.8$ and $N$ changes.
The frequencies of vibration are marked by the yellow bar. Dashed lines are fits of the
Eq.~\eqref{sp_dhc} to the experimental data in regions II and III.
Thick-dashed lines show the limit behaviors $f^{-2}$ and $f^{-1}$. 
\label{fig:spettri}}
\end{figure}

{\em Velocity power density spectrum (vpds)}: in
Fig.~\ref{fig:spettri}, we present our main results in the form of the
power density spectrum of the velocity signal $\omega(t)$, which is
defined as $S(f)=\frac{1}{2\pi t_{TOT}}\lvert\int_0^{t_{TOT}}
\omega(t) e^{i (2\pi f) t}dt \rvert^2 $. We recognize $4$ frequency
ranges, denoted as regions I, II, III and IV. In the experiments at
fixed (maximum) density, $N=2600$, displayed in
Fig.~\ref{fig:spettri}A, the spectrum conserves the same qualitative
shape, vertically shifted due to the differences in $\Gamma$. The most
striking properties are observed in regions II and III: $S(f)$ goes
from a plateau in region II to a roughly parabolic maximum centered at
a frequency $f^*$ in region III. The value of the resonant frequency
$f^*$ slightly shifts from $10$ Hz to $20$ Hz as $\Gamma$
decreases. To avoid interference, we ensured that the shaker vibration
is in a distant region ($200-400$ Hz); we changed such a range
(including trials with a single frequency, i.e. a harmonic vibration),
obtaining always the same shape $S(f)$ with same values of $f^*$. A
mechanical resonance is also observed at $\sim 70$ Hz, due to the
non-perfect acoustic insulation of the plate on which the couple
encoder/blade is mounted. In conclusion, the maximum in $f^*$ is a
resonance experienced by the blade in its motion through the granular
medium, and we interpret it as a transient trapping phenomenon,
analogous to caging effects in low-temperature/high-density
liquids. We will see that such an interpretation is well supported by
other observations and by a theoretical model. In both ``extremal''
regions I and IV, $S(f)$ is a decreasing function of $f$. In
particular, in region IV, the high-frequency decay presents a power
law $\sim f^{-\beta}$ with $1 < \beta < 2$, close to $1$ for the
lowest values of $\Gamma$. This is the evidence of collective effects,
in the fast processes {\em inside a cage} (as $f \gg f^*$), occurring without a
characteristic frequency~\cite{unosuf}. The decay in region I is also
anomalous and denotes the emergence of long characteristic times,
possibly larger than the experiment duration $t_{TOT}=3600$ s. Note that the asymptotic diffusion is governed by $\lim_{t
\to \infty} \langle \Delta \theta^2(t) \rangle/t \sim 2\pi S(f \to 0)
$: therefore an increasing value of $S(f)$ as $f\to 0$ (i.e. at
increasing time) indicates a superdiffusive behavior, detailed below.

When the density is reduced, see Fig.~\ref{fig:spettri}B, the shape of
$S(f)$ drastically changes. The slope of the decay in region I
decreases and eventually vanishes: at low densities a plateau spans
both regions I and II. The maximum in region III is reduced and
disappears for $N<1000$, called the ``gas''-phase. The exponent of the
power-law decay in region IV increases $\beta \to 2$. The whole
spectrum at the lowest densities is well fit by the Lorentzian
$S(f)=\frac{T}{\pi \gamma}/[1+(2\pi If/\gamma)^2]$, expected for
diffusion in diluted gases at temperature $T$ with a collision
frequency $\propto \gamma$~\cite{gasdiff}. Here $T$ is the probe's ``kinetic
temperature'' $T=I\langle\omega^2 \rangle$.

\begin{figure}
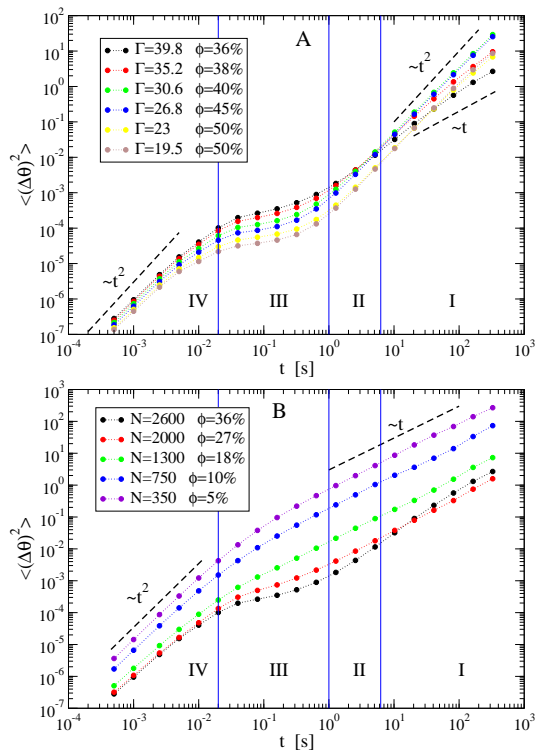

\includegraphics[width=7cm,clip=true]{diffusione_varioA}
\includegraphics[width=7cm,clip=true]{diffusione_varioN}
\caption{
Mean-squared displacement $\langle [\Delta \theta(t)]^2\rangle$ after
a time $t$ for the two series. A: $N$ = 2600 and $\Gamma$ varies. B:
the shaking intensity is fixed to $\Gamma$=39.8 and $N$ changes. Black
dashed lines are guides for the eye.
\label{fig:msd}}
\end{figure}

{\em Mean-square-displacement (msd)}. The several phenomena observed
in $S(f)$ are reflected in the diffusion properties, see
Fig.~\ref{fig:msd}, where the mean-square displacement $\langle
[\Delta \theta(t)]^2\rangle$ after a time $t$ is plotted. The four
temporal regions corresponding to the frequency regions discussed
above are marked on the graph. Our ``cage-like'' interpretation of the
maximum of $S(f)$ in region III is corroborated, in
Fig.~\ref{fig:msd}A, by the dramatic slowing-down of $\langle [\Delta
  \theta(t)]^2\rangle$ in the same region~\cite{reis2007}, resembling
the typical dynamical slowdown in the diffusion of tracers dispersed
in viscous liquids. At small times (region IV) the usual ballistic
behavior appears. More remarkable is the behavior at large times. All
the experiments with $N=2600$ present a superdiffusive range $\sim
t^\alpha$ in region II, with $\alpha>1$. For the largest values of
$\Gamma$, this behavior changes to a diffusive behavior $\sim t$ in
region I. On the contrary, at lower $\Gamma$, the superdiffusive
exponent $\alpha>1$ remains the same of region II at large times. In
particular, for $\Gamma < 31$, we find an almost ballistic
superdiffusion $\alpha \sim 2$. The situation is very different when
the density is reduced (Fig.~\ref{fig:msd}B): the long time behavior
(region I) is always of the normal type $\sim t$. In the most dilute
cases ($N<1000$) the typical scenario of diffusion in a gas-like fluid
is fully recovered in the form of a monotonic crossover from the
ballistic region IV to the normal diffusion of region II and I. 
Changing the size and shape of the blade, see~\cite{sm}, does not lead to relevant changes in
the above scenario.

\begin{figure}
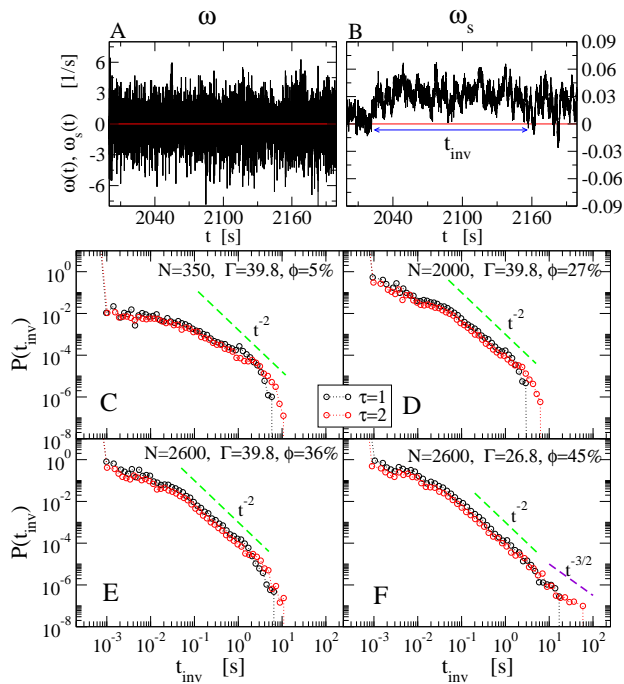

\hspace{.8cm}\includegraphics[width=7.5cm,clip=true]{decomposition}
\includegraphics[width=8cm,clip=true]{persistency}
\caption{A and B: angular velocity $\omega(t)$ and filtered
  signal $\omega_s(t)$ (with a running average over a time $\tau=2$
  s), for $N=2600$ and $\Gamma=26.8$. C,D,E,F: pdf of the inversion
  times calculated (with $\tau=1$ s and $\tau=2$ s) in different
  experiments. From C to F the following configurations are displayed:
  very dilute, intermediate, dense at high energy, dense at low
  energy. Colored dashed lines are guides for the eye.
\label{fig:ctrw}}
\end{figure}

The study of vpds and msd are consistent with measurements of the
velocity-autocorrelation function (vacf) $C(t)$, which is the inverse
Fourier transform of $S(f)$. In the very dilute cases, it is close to
a simple exponential decay.  The most dense and ``cold'' experiments,
on the contrary, reveal a vacf with many features: a fast, though
non-exponential, decay at small times, followed by a back-scattering
oscillation through negative values, interpreted as the ``cage'', and
finally a slow decay to zero~\cite{zippelius}. The final decay of the
vacf could also light up the origin of superdiffusion. Unfortunately
at large times the vacf is exceedingly noisy. A more promising way to
probe long memory effects is to measure the times of persistency,
i.e. the times during which the signal remains positively
correlated. Our operative definition consists in two basic steps: 1)
filter out high frequency oscillations which are not relevant for the
behavior at large times of $\Delta \theta(t)$, by taking the running
average $\omega_s(t)=\frac{1}{\tau}\int_t^{t+\tau} \omega(t')dt'$ over
a large time $\tau \ge 1s$; and then 2) compute the statistics of the
times separating two consecutive zeros of $\omega_s(t)$, which we call
inversion time $t_{inv}$, see Fig.~\ref{fig:ctrw}A and~\ref{fig:ctrw}B
for experimental samples of $\omega(t)$, $\omega_s(t)$ and $t_{inv}$.
The statistics of $t_{inv}$ is a natural measurement of long term
memory of the signal. The experimental probability density (pdf) of
$t_{inv}$ is shown in Fig.~\ref{fig:ctrw}C-~\ref{fig:ctrw}F for a few choices of
parameters and values of $\tau$. We observe that it rapidly decays
with an exponential cut-off smaller or equal than $\sim 10 s$ in all
cases where the msd asymptotically showed normal diffusion, signaling
a finite memory of the dynamics. The cut-off apparently jumps to a
much larger value in the cases where the msd displays superdiffusion:
in such cases, the pdf decays to zero as a power $\sim -2$ or even
slower. This is a fair evidence that long memory effects arise
together with the observed superdiffusion. Long memory may be due to a
slow rotating creeping motion of the surrounding granular medium,
which acts as a coherent block and drags the blade. Such a
``secondary'' motion has long relaxation times due to the involved
large inertia. New experiments are being designed with the aim of
demonstrating this picture.  We mention that longer experiments (shown
in the SM~\cite{sm}) still display superdiffusion that evolves in
normal diffusion after many hours.

 {\em Modelling the dynamics of the blade}. In the most dense and
  cold experiments, the vpds and the msd display an intriguing
  superposition of phenomena over almost $7$ decades of
  time-scales. It seems hopeless to reduce such a complexity to a
  model with a simple physical interpretation. It is tempting to
  disentangle the two main phenomena, i.e. caging (occurring at times
  smaller than $10^{-1} s$) and superdiffusion (more evident at times
  larger than $10 s$), by decomposing the dynamics into a slow and a
  fast component $\omega(t)=\omega_s(t) + \omega_f(t)$, with
  $\omega_s(t)$ defined above. Experimental data demonstrate that the
  standard deviation of $\omega_f(t)$ is much larger than that of
  $\omega_s$ and therefore dominates at short times (large
  frequencies).  At long times (small frequencies), on the contrary,
  the fast dynamics averages out and $\omega_s(t)$ emerges
  as the leading signal.  For the fast dynamics, we propose a simple
  interpretation of the transient caging phenomena through a
  ``diffusing harmonic cage'' (dhc) model: this is a simplified
  version of the Itinerant Oscillator model describing translational
  and rotational diffusion of particles in liquids~\cite{io1,io2}, and
  reads
\begin{subequations} \label{dhc}
\begin{align}
\dot{\theta}(t)&=\omega_f(t) \;\;\;\;\;\;\;\;\;\;\; \dot{\theta}_0(t)=\sqrt{2D_0}\xi'(t)\\
I\dot{\omega}_f(t)&=-\gamma\omega_f(t)-K[\theta(t)-\theta_0(t)]+\sqrt{2 \gamma T}\xi(t),
\end{align}
\end{subequations}
where $\xi(t)$ and $\xi'(t)$ are white normal Gaussian noises. The
model represents the diffusion of a particle in a harmonic potential
with ``stiffness'' $K$ and unfixed minimum, under the effect of a
thermal bath at temperature $T$ and relaxation time $I/\gamma$. The
harmonic potential, representing the cage due to the confining effect
of the dense granular host fluid, is not fixed but moves, as
$\theta_0(t)$ behaves as Brownian motion with diffusivity
$D_0$. Motivation for this model is twofold: 1) the main features of
the vpds, i.e. an elastic resonance (region III) and a plateau
revealing loss of memory at larger times (region II); 2) in the dilute
limit it can be rigorously derived~\cite{bromo}, while at intermediate
densities a series of studies showed that memory effects (coming from
correlated collisions) are well described by a coupling with an
additional degree of freedom fluctuating at slower
time-scales~\cite{gasdiff}.  The vpds of the above model can be
calculated and reads
\begin{equation} \label{sp_dhc}
S(f) = \frac{1}{\pi}\frac{D_0 K^2+\gamma T (2\pi f)^2}{\gamma^2 (2\pi f)^2+[K-I(2\pi f)^2]^2}~~.
\end{equation}
Two limiting cases are recovered: when $K=0$, the Ornstein-Uhlenbeck
process is obtained, with $S(f)$ taking the Lorentzian form mentioned
before. When $K>0$ and $D_0=0$, one has the Klein-Kramers process in a
fixed harmonic potential, and $S(f) \to 0$ for $f \to 0$, expressing
the absence of diffusion at large times: the cage does not move and
fully confines the particle.  Formula~\eqref{sp_dhc} fairly fits all
experimental spectra in regions II and III, with parameters given in
Tables~2 and~3 of~\cite{sm}.  Reasonably, the ``cage stiffness''
decreases at increasing shaking intensity. It also decreases as the
density is reduced, and abruptly goes to zero for $N\sim 1000$. The
``cage diffusivity'' $D_0$ rapidly increases with increasing $\Gamma$
and with decreasing $N$. A more detailed study of the transition from
the cage behavior $K\neq 0$ to the free behavior $K=0$ is postponed to
future investigations.

 A more ambitious task is to devise a simple mechanism for
 $\omega_s(t)$, leading to superdiffusion. In driven granular systems
 it has been observed below the jamming transition~\cite{bouchaud} and
 in a few cases above it, where it was imputed to ``zero''-modes of
 the host fluid~\cite{barrat}, to Taylor dispersion~\cite{behringer},
 or to turbulence-like cascade effects~\cite{roux}. We stress that, at
 variance with standard diffusion, for anomalous diffusion there is
 nothing similar to universality~\cite{klages}. A systematic
 derivation of anomalous diffusion is a hard task and it is possible
 only in few specific cases~\cite{klages}. Of course, the basic
 ingredient must be an enduring memory: a way to achieve it is to
 replace standard time derivatives with fractional derivatives in the
 Fokker-Planck equations, an approach which is subject of a vast
 literature~\cite{klafter}. Such an approach is capable, in principle,
 of describing both caging and superdiffusion within a single model
 equation, at the price of losing immediate interpretation and plain
 calculations.  In a complementary approach~\cite{angelo} a
 coarse-grained value of $\omega_s(t)$ (where only the sign of this
 quantity is traced) follows a continuous time random walk (ctrw): It
 takes discrete values with random transition times extracted from a
 given distribution. A simplified version of the ctrw is discussed in
 the SM~\cite{sm}. We highlight that the ctrw gets a suggestive
 experimental confirmation in the observations of Fig.~\ref{fig:ctrw}:
 indeed the ctrw model quantitatively connects the observed slow decay
 of $P(t_{inv})$ to superdiffusion.

{\em Conclusion}: an experimental study of dense granular mesorheology
allows to probe time-scales in a range of six orders of
magnitude. Such an investigation reveals a complex scenario with
different dynamical behaviors in four frequency or time regions. Three
crucial features are observed at large density and low temperature: a
``resonant'' caging phenomenon at intermediate scales, non-white
noise at fast scales, and super-diffusion at long times.  The caging
phenomenon is compatible with a diffusing harmonic cage model, while
superdiffusion seems to be rooted in the long inversion times
appearing in the dynamics $\omega_s(t)$, possibly related to creeping
rotating motion of the granular media. A more detailed investigation
of {\em transitions} is in order: a gas-liquid transition could be put
in evidence by studying how $K\to 0$ when $N$ decreases; a
liquid-glass transition could be taking place at the lowest values of
$\Gamma$ or, eventually increasing $N>2600$. Superdiffusion and large
values of $t_{inv}$ could be signaling such a transition. Another
promising line of investigation is {\em active} mesorheology,
i.e. probing the response of the system to the application of an
external force, achieved by coupling a controllable motor to the
blade~\cite{noiplos}.

\begin{acknowledgments}
We wish to thank A. Lasanta Becerra and A. Sarracino for useful
discussions.  AP acknowledges the support of the Italian MIUR
grants FIRB-IDEAS n. RBID08Z9JE. 
\end{acknowledgments}

%%%%%%%%%%%%%%%%%%%%%%%%%%%%%%%%%%%%%%%%%%%%%%%%%%%%%%%%%%%%%%%%%%%%%%%%%%%%%%%%%%%
%%%%%%%%%%%%%%%%%%%%%%%%%%%%%%%%%%%%%%%%%%%%%%%%%%%%%%%%%%%%%%%%%%%%%%%%%%%%%%%%%%%
%%%%%%%%%%%%%%%%%%%%%%%%%%%%%%%%%%%%%%%%%%%%%%%%%%%%%%%%%%%%%%%%%%%%%%%%%%%%%%%%%%%

%Previous studies SUPPORTING OUR RESULTS:
%
%\begin{itemize}
%
%\item ``micro''-super-diffusion in experiments above the jamming transition, in~\cite{bouchaud} (experimental vibrated dense monolayer)
%
%\item super-diffusion in simulation just below the jamming transition, in~\cite{barrat}, discussion of ``zero''-modes of the host fluid (simulated quasi-static shear)
%
%\item super-diffusion due to ``taylor dispersion'' along the flow direction, in~\cite{behringer} (experimental couette shear)
%
%\item turbulent-like super-diffusion ($5/4$ power spectrum, long time memory) in~\cite{roux} (simulated quasi-static shear)
%
%\item cage effects and vacf with long tail , in~\cite{zippelius} (simulated random driving)
%
%\end{itemize}

%\vspace{1cm}

\newpage
\begin{widetext}

{\bf SM: SUPPLEMENTAL MATERIAL}

%\vspace{1cm}
\section{Details of the experimental setup}

The scheme of our experimental setup is shown in
Fig.~1A. A Poly(methyl methacrylate) (PMMA) container
(diameter 90 mm, maximum height 47 mm and total volume 245 cm$^{3}$)
with an inverse-conical-shaped base, is filled with $N$ steel spheres
(diameter 4 mm) and mounted on an electrodynamic shaker (LDS V450)
which is fed, through an amplifier, with a noisy signal whose spectrum
is approximately flat in a range $[f_{min},f_{max}]$ with
$f_{min}=200$ Hz and $f_{max}=400$ Hz. The shaker responds roughly
linearly in the same range of frequencies, as verified with an
accelerometer placed on the container. The elevation of the container
at time $t$ is denoted by $z(t)$. We give the range of values for the
relevant variables: for the maximum acceleration $\ddot{z}_{max} \sim
20-40$ g, the maximum velocity $\dot{z}_{max} \sim 80-300$ mm/s and
the maximum displacement $z_{max} \sim 0.03-0.25$ mm. Details are
reported below in Table~1. A PMMA blade (dimensions
$35\times6\times15$ mm, momentum of inertia $I=353$
g$\cdotp$mm{$^2$}), suspended to a mounting which is mechanically
isolated from the container/shaker apparatus, is immersed into the
granular medium and can rotate around a vertical axis placed at the
center of the container. The angular velocity $\omega(t)$ of the blade
and its absolute (i.e. without the modulus $2\pi$ operation) angle of
rotation $\theta(t)=\int_0^t \omega(t')dt'$ are measured by an angular
encoder (AEDA-3300) at a time-resolution of $2$ kHz and angular
resolution 40000 divisions per revolution. The granular medium is in a
liquid state because of the vibration. The blade, under the effect of
the interactions with the spheres, performs a motion qualitatively
similar to an angular Brownian motion. It is important to remark that
the blade can only rotate and therefore moves along the ``free flow''
direction, i.e. it only probes the fluctuations of granular modes
parallel to the boundaries of the container. Moreover, the blade's
size is such that, in the dense cases, its rotation is only allowed by
the rearrangement of a large number of surrounding particles. We have
performed two families of experiments: experiments at high density
($N=2600$) and different normalized (i.e. in $g$ units) shaking
intensities $\Gamma=\ddot{z}_{max}/g \in[19.5,39.8]$, and experiments
at high shaking intensity ($\Gamma=39.8$) and different densities $N
\in [300,2600]$. In Fig.~1B and C we have reported the
values of the mean squared angular velocity $\langle \omega^2 \rangle$
of the blade in the different experiments.

%%%%%
\begin{table*}[hpb]
\begin{center}
\begin{tabular}{| c | c | c | c | c | c | c|}
  \hline
  $\Gamma$ & 39.8 & 35.2&30.6& 26.8&23&19.5 \\
    \hline
  $ z_{min} - z_{max}$ & 0.25-0.06 & 0.22-0.05 & 0.19-0.05 & 0.17-0.04&0.14-0.04&0.12-0.03 \\ 
    \hline
  $ \dot{z}_{min} - \dot{z}_{max}$ &  310.4-155.2 & 274.5-137.3 &238.6-119.3& 209-105 & 179.4-89.7 &152-76 \\
  \hline
\end{tabular}
\end{center}
\caption{Equivalence between the normalized shaking intensity $\Gamma$ and the physical amplitudes and velocities of the container, expressed in mm and mm.s$^{-1}$, with $f_{min}$ = 200 Hz and $f_{max}$ = 400 Hz \label{tab:shaker}}

\end{table*}
%%%%%

\section{Fits of the experimental data: parameters}

%%%%
\begin{table}[htb]
\begin{center}
\begin{tabular}{|c|c|c|c|c|c|c|}
\hline
$\Gamma$ & $D_0$    &  $K/[I (2\pi)^2]$          & $\gamma/I$      &    $T/I$    & $\frac{T/I}{\langle \omega^2\rangle}$  & $T/\gamma$ \\ 
  $g$  & $10^{-5}~s^{-1}$      &  $s^{-2}$        & $s^{-1}$           &    $s^{-2}$  &                                        & $s^{-1}$ \\
\hline
 39.8	&$46~\pm	~4$		&$128~\pm~3	$&$121~\pm~3		$& $0.51~\pm~  0.01 	$&0.54      	& 0.0042  \\
35.2	&$31~\pm	~3$		&$151~\pm~4	$&$134~\pm~4		$&$0.47~\pm~   0.01	$&0.53     	& 0.0035 \\
30.6	&$13~\pm	~2$		&$181~\pm~4	$&$151~\pm~4		$&$ 0.371~\pm~  0.009 	$&0.51     	& 0.0024 \\
26.8	& $8  ~\pm~1$		&$215~\pm~5	$&$171~\pm~5		$&$0.307~\pm~   0.007 	$&0.5      	& 0.0018 \\
23	& $4.6 ~\pm~0.8$	&$260~\pm~7	$&$217~\pm~7		$&$0.241~\pm~ 0.006  	$&0.47     	& 0.0011 \\
19.5	& $2.6 ~\pm~	0.5$	&$320~\pm~7	$&$250~\pm~7		$&$0.198~\pm~   0.005	$&0.44     	& 0.00079 \\
\hline
\end{tabular}
\caption{Fit's parameters with Eq.~2, series $N=2600$ and $\Gamma \in$ [19.5, 39.8]. \label{tab:fit1}}
\end{center}
\end{table}
%%%%

%%%%
\begin{table}[htb]
\begin{center}
\begin{tabular}{|c|c|c|c|c|c|c|}
\hline
$N$     & $D_0$         &  $K/[I (2\pi)^2]$        & $\gamma/I$    &   $T/I$     & $\frac{T/I}{\langle \omega^2\rangle}$     & $T/\gamma$ \\ 
   & $10^{-3}~s^{-1}$      &  $s^{-2}$        & $s^{-1}$           &    $s^{-2}$  &                                        & $s^{-1}$ \\
\hline
2600		&$0.46~\pm~0.04$		&$128~\pm~3	$&$121~\pm~3		$& $0.51~\pm~  0.01 	$&0.54      	& 0.0042  \\
2000		&$1.44~\pm~0.04$		&$88~\pm~2	$&$103~\pm~2		$&$0.57~\pm~   0.02	$&0.52     	& 0.0055 \\
1300		&$7.7~\pm	~0.1$	&$84~\pm~18	$&$131~\pm~6		$&$ 0.97~\pm~  0.08 	$&0.54     	& 0.0074 \\
700 (gas)	& 0					&$0			$&$102~\pm~9		$&$7.0~\pm~   1. 	$&1.1      	& 0.069 \\
350 (gas)	& 0					&$0			$&$79~\pm~16		$&$16.0~\pm~   2.	$&1.1     	& 0.2 \\
  \hline
\end{tabular}
\caption{Fit's parameters with Eq.~2, series $\Gamma=39.8$ g and $N \in$ [350, 2600].  \label{tab:fit2}}
\end{center}
\end{table}
%%%%

\section{Experiments with blades of different sizes}

We have performed a number of experiments with different blades,
changing the total scattering surface and its shape. The
thickness of the blades being constant, as well as its material, the mass and
momentum of inertia are also varied accordingly.

In Figure~\ref{sizes}{\em a} we portray the four shapes used. Shape
``A'' is the one used in the rest of the paper. Panels {\em b} and
{\em c} of Figure~\ref{sizes} represent the vpds and the msd
respectively, with the four shapes, in the case with $\Gamma=30.6$ and
$N=2600$, corresponding to packing fraction $\phi=40\%$. It appears
that the general features of the vpds and msd are conserved when the
size and shape of the blade are changed. Higher or smaller energy is
obviously observed, when the inertia is reduced (shapes C and D) or
increased (shape B). Differences in the mechanical (spurious)
resonance at $\sim 70-100$ Hz are due to some modifications of the
mounting mechanism on which the blade is fixed, occurred between the
old and the new experiments. In the msd of the larger balde (shape B)
it is observed - at large times - a saturation of the mean squared
displacement which we impute to border effects. Note also that at very
large time the measurement of the msd is noisy and prone to errors.

\begin{figure}[hb]
\includegraphics[height=5.2cm,clip=true]{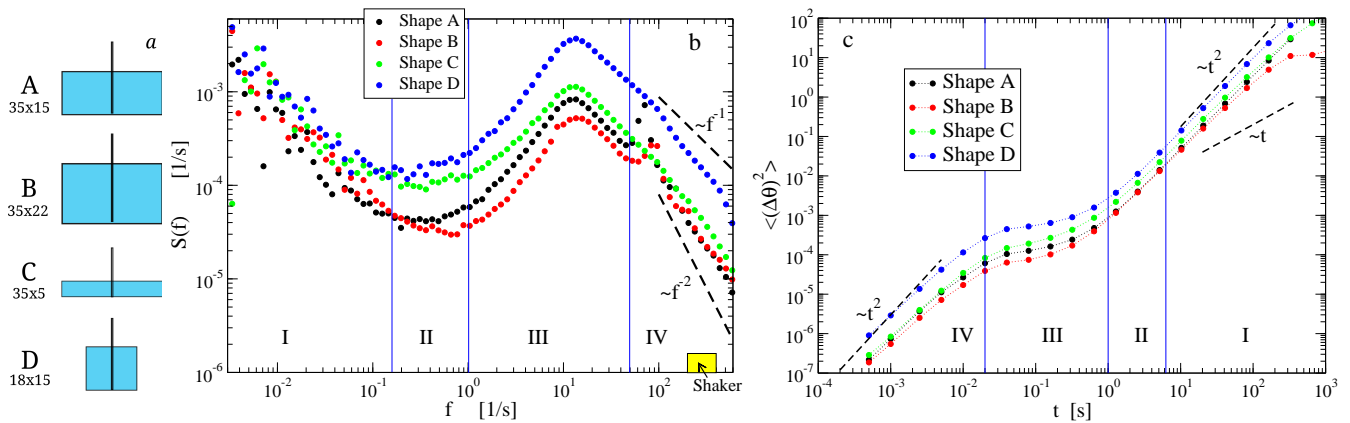}
\includegraphics[height=5.5cm,clip=true]{spettri_varioSIZE.eps}
\includegraphics[height=5.5cm,clip=true]{diffusione_varioSIZE.eps}
\caption{Effect of the size and shape of the blade. The four shapes are represented in panel {\em a}. In panel {\em b} we show the velocity power density spectrum for the four cases, with $\Gamma=30.6$ and $N=2600$ which corresponds to packing fraction $\phi=40\%$. For the same parameters, the behavior of the mean squared displacement is displayed in panel {\em c}. \label{sizes}}
\end{figure}

\section{Very long experiments}

The finite size of the experiment suggests that all relaxation times,
even if long, are finite: the displacement $\Delta \theta$, therefore,
is expected to reach asymptotically a normal diffusive behavior. We
have performed a $12$-hours-long experiment in order to check such a
hypothesis. The mean squared displacement is shown in
Fig.~\ref{fig:longmsd}. Indeed, at large times the
superdiffusive behavior shows a crossover toward normal diffusion, which is slowly approached after many hours.

\begin{figure}[h]
\includegraphics[height=5.2cm,clip=true]{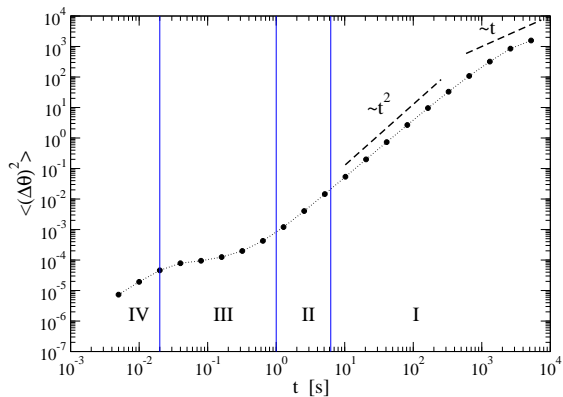}
\caption{Mean squared displacement in a long experiment (total time $12$ hours), with $N=2600$ and $\Gamma=26.8$. \label{fig:longmsd}}
\end{figure}

\section{The continuous time random walk model}

A possible mechanism for superdiffusion is the {\em continuous time
  random walk} model (ctrw)~\cite{angelo,klages}. In the ctrw the
variable (here the velocity) performs a random walk taking discrete values $\omega_i$: it
remains constant for a random time $t$ and then jumps to a new
value. The choice of the statistics (pdf) of $t$ is the main parameter
of the model and determines the properties of the autocorrelation of
the velocity and, therefore, of the mean squared displacement. 

Here we consider in details a simplified version of the ctrw model,
where the velocity can take only two values. Such a simplification does not change the substance of
the demonstration, which can be easily generalised to models with many
possible values.  Note that a comparison between this model and
the experimental results should be made only at long time
scales (i.e. such that $\Delta \theta(t)$ is well approximated by $\int_0^t\omega_s(s)ds$) and in the spirit of a strong
coarse-graining, as if a dramatic loss of resolution allows to measure only the sign of the velocity.

In the model, $\omega_s$ changes its value at
times $t_n$ with $n \in [0,m]$ and $t_m=t$. In the interval between
$t_{n-1}$ and $t_n$, $\omega_s$ takes the value $\omega_{0,n} \in [-\omega_0,\omega_0]$. The
intervals $t_{inv,n}=t_{n}-t_{n-1}$, with $\sum_1^{m} t_{inv,n} = t$,
are random and independent with the following probability density
function
\begin{equation}
P(t_{inv}=x) \sim \left\{\begin{matrix} g(x) \;\;\;\;&x \in[0,t^*]\\ x^{-g} \;\;\;\;&x \in(t^*,t_{max}]\\ 0 \;\;\;\;&x>t_{max}. \end{matrix} \right.,
\end{equation}
where $g(x)$ is a smooth function: its shape and the value of $t^*$ (assumed to be always finite) are not relevant for the long time behavior of the displacement.
The cut-off $t_{max}$ may
eventually be taken to $\infty$ (see below). 

For the msd  therefore we have
\begin{equation} 
\langle [\Delta \theta(t)]^2\rangle = \left\langle \left[\int_0^t \omega_s(s)ds \right]^2 \right\rangle = \omega_0^2 \sum_{n=1}^{m} \langle t_{inv,n}^2 \rangle + \sum_{n\neq k} \langle \omega_{0,n}\omega_{0,k} \rangle \langle t_{inv,n} t_{inv,k}\rangle
\end{equation}
and the values $\omega_{0,n}$ being independent and with zero average, one may drop the last term obtaining
\begin{equation} 
\langle [\Delta \theta(t)]^2\rangle = \omega_0^2 \sum_1^{m} \langle t_{inv,n}^2 \rangle= m \omega_0^2 \langle t_{inv,n}^2 \rangle
\end{equation}
The presence of the cut-off allows us to write
\begin{equation}
\langle t_{inv} \rangle = \frac{t}{m}.
\end{equation}
and therefore one finally has 
\begin{equation} \label{ctrw_msd}
\langle [\Delta \theta(t)]^2\rangle = \omega_0^2\frac{\langle t_{inv}^2\rangle}{\langle t_{inv} \rangle}t ~~.
\end{equation}

The dependence on the cut-off (at large
values of $t_{max}$) of the moments of order $k\ge 1$ of $P(t_{inv})$
are determined by the distribution at times larger than $t^*$, i.e.
\begin{equation}
\lim_{t_{max} \to \infty} \langle t_{inv}^k \rangle \sim \lim_{t_{max} \to \infty}  \int_{t^*}^{t_{max}} x^{k-g} dx \sim \left\{ \begin{matrix} \textrm{const.} \;\;\;\; &g>k+1\\ t_{max}^{k+1-g} \;\;\;\; &g\le k+1~, \end{matrix}\right.
\end{equation}
The last equality, together with Eq.~\eqref{ctrw_msd}, gives three possible behaviors for the
msd at large times $t \gtrsim t_{max}$:
\begin{itemize}

\item for $g>3$: normal diffusion, independently of the cut-off
  $t_{max}$;

\item for $2<g\le 3$: $ \langle [\Delta \theta(t)]^2\rangle \sim t_{max}^{3-g}\;\; t$ ;

\item  for $1<g\le 2$:
$\langle [\Delta \theta(t)]^2\rangle \sim t_{max}\;\; t$.

\end{itemize}
For $1<g \le 3$, in the range $t \leq t_{max}$, one expects $\langle
[\Delta \theta(t)]^2\rangle \sim t^{\beta}$. The value of $\beta$ can
be determined by asking that it matches, at $t \sim t_{max}$, the
asymptotic behavior given above: for instance in the case $2<g\le 3$
the behaviors $\langle [\Delta \theta(t)]^2\rangle \sim
t_{max}^{3-g}\;\; t$ and $\langle [\Delta \theta(t)]^2\rangle \sim
t^{\beta}$ can match only if $\beta=4-g$; the same argument for
$1<g\le 2$ gives $\beta=2$. The situation is
therefore summarized here
\begin{equation} \label{ctrwfinal}
\langle [\Delta \theta(t)]^2\rangle \sim \left\{ \begin{matrix} t \;\;\;\;&g>3\;\;\textrm{or}\;t \gtrsim t_{max}\\ \;\;\;\; t^{4-g} \;\;\;\;&2<g \le 3 \;\;\textrm{and}\;t < t_{max} \\ t^2 \;\;\;\;&1<g \le 2\;\;\textrm{and}\;t<t_{max}.\end{matrix} \right.
\end{equation}
 Fig.~4 of the Letter shows that $P(t_{inv})$ has an  exponential cut-off at
 $t_{max}\lesssim 10$ seconds for all the experiments where diffusive
behavior is observed, that is in the most dilute or energetic
cases. The cut-off rapidly grows and reaches times of the order of the duration of the experiment when $N=2600$
and $\Gamma < 31$: a power law tail with $g \lesssim 2$
emerges. Equation~\eqref{ctrwfinal} makes this power law decay fully consistent with the ballistic behavior observed
in Fig.~3 of the Letter.

\end{widetext}

\end{document}